\documentclass[english,keywords,aps,amsmath,amssymb,twocolumn]{revtex4-1}
\usepackage[T1]{fontenc}
\usepackage[latin1]{inputenc}
\usepackage{babel}
\usepackage{amssymb}
\usepackage{graphicx}
\usepackage{color}
\usepackage{bm}
\usepackage{longtable}
\usepackage{amsmath} 
\usepackage{amsfonts}
\usepackage{amssymb}
\begin{document}

\title{ Inequivalent Leggett-Garg inequalities}

\author{Swati Kumari }
\author{A. K. Pan \footnote{akp@nitp.ac.in}}
\affiliation{National Institute of Technology Patna, Ashok Rajpath, Patna, Bihar 800005, India}
\begin{abstract}
It remains an open question how realist view of macroscopic world emerges from quantum formalism. For testing the macrorealism in quantum domain, an interesting approach was put forward by Leggett and Garg in $1985$, by formulating a suitable  inequality valid for any macrorealistic theory. Recently, by following the Wigner idea of local realist inequality, a probabilistic version of standard Leggett-Garg inequalities have also been proposed. While the Wigner form of local realist inequalities are equivalent to the two-party, two-measurements and two outcomes CHSH inequalities, in this paper we provide a generic proof to demonstrate that the Wigner form of Leggett-Garg inequalities are \textit{not only} inequivalent to the standard ones, but also stronger than the later. This is demonstrated by quantifying the amount of  disturbance caused by a prior measurement to the subsequent measurements. In this connection, the relation between LGIs and another formulation of macrorealism known as no-signaling in time is examined.
\end{abstract}

\maketitle

\section{Introduction}
Bell's  inequalities \cite{bell} were formulated for testing the incompatibility between the local realism and quantum mechanics (QM). Local realism is the notion that objects have definite properties independent of the observation, and measurements of these properties are not affected by  space-like separated events. Schrodinger's \cite{sch} famous cat experiment had raised the question how the notion of macrorealism inherent in our everyday experience can be reconciled with the quantum formalism. Since then, quite a number of attempts have been made to answer this question \cite{arndt,zeh,ghi,bruk}. Motivated by the Bell's theorem \cite{bell}, in 1985, Leggett and Garg \cite{leggett85} formulated an inequality which provides an elegant scheme for empirically testing the incompatibility between the classical world view of macrorealism and quantum mechanics. 

The notion of macrorealism consists of two main assumptions \cite{leggett85,leggett,A.leggett} which are in principle valid in our everyday world are the following;
	
	\emph{ Macrorealism per se :} If a macroscopic system has two or more macroscopically distinguishable real states available to it, then the system remains in one of those states at all instant of time.
	
	\emph{Non-invasive measurability :} The definite real state of the macrosystem is determined without affecting the state itself or its possible subsequent dynamics.
	
 Based on these assumptions, the standard Leggett-Garg inequalities (henceforth, SLGIs) were derived. Such inequalities can be shown to be violated in certain circumstances, thereby implying that either or both the assumptions of macrorealism \emph{per se} and non-invasive measurability are not compatible with all the quantum statistics. Since then flurry of theoretical proposal \cite{budroni, budroni15,maroney,kofler,clemente,UshaDevi,emary,saha,moreira,clemente16,hall,mal} have been given and quite a number of experiments \cite{a.p,goggin,xu,dressel,suzuki,julsgaard,isart,mahesh11,souza,katiyar13,formaggio,katiyar}  have also been performed. However, there is active debate \cite{maroney,hall} how non-invasive measurability can be ensured within the framework of quantum mechanics and in real experiments. Besides SLGIs, there have been other interesting  formulations for testing the macrorealism, such as, Wigner form of Leggett-Garg inequalities (WLGIs)\cite {saha} and no-signaling in time condition \cite{clemente,clemente16}. 

Fine\cite{fine} showed that for a two-qubit system subject to two measurements per qubit having two outcomes of each measurement,  the only relevant Bell's inequality is the CHSH form \cite{chsh}. Any other form of inequalities reduce to the CHSH inequality. However, if the number of measurements and/or system size  is more than two, the inequalities inequivalent to the CHSH type inequality can be found \cite{cglmp,col}. Then CH form \cite{ch} and Wigner form \cite{ep,cas,home15} of local realistic inequalities are equivalent to CHSH inequality. 

SLGIs are often considered to be the analogus to the CHSH inequalities. But, this structural analogy is only cosmetic. It is recently shown by Budroni {et al.}\cite{budroni} that  the violation of a SLGI can reach up to its algebraic maximum while violation of a CHSH inequality is contrained by Tsirelson bound. It is more crucial to note here that for two-party, two-measurements and two outcomes Bell scenario (known as, 2222 Bell scenario), while CHSH inequalities provide the necessary and sufficient condition for local realism \cite{fine}, in a recent paper, Clemente and Kofler \cite{clemente16} have argued that \emph{no} set of SLGIs can provide the same for macrorealism. However, they argued that a suitable conjunction of two-time and three-time no-signaling in time conditions provide necessary and sufficient condition for macrorealism. In terms of a macrorealism polytope, it is shown \cite{clemente16} that SLGIs does not represent the facets of that polytope, rather it is a hyperplane. That opens up the possibility that a different set of inequalities (say, WLGIs) can be inequivalent to and stronger than SLGIs. 

In this paper, we provide a generic proof to show that WLGIs are \textit{not} only \textit{inequivalent} to the SLGIs but also \textit{stronger} than SLGIs. In other words, we show that if any of the possible symmetries of SLGIs is violated by QM then one can find at least a WLGI which is also violated but converse is \textit{not} true. In this connection, by adopting a different line of argument than in \cite{clemente}, we demonstrate here that how no-signaling in time conditions capture the macrorealism better than WLGIs and consequently SLGIs. In order to showing this, we invoke the notion of the amount of disturbance created by the prior measurement to the subsequent measurements. We first derive disturbance inequalities corresponding to LGIs and WLGIs which enable us to provide the aforementioned  proof.  We further pointed out that the violation of LGIs and WLGIs require a threshold value of disturbance and the suitable interplay between them, but any non-zero value of disturbance implies the violation of no-signaling in time condition, thereby making it a better candidate for testing macrorealism.

This paper is organized as follows. In Section II, we recapitulate the notions of  SLGIs and WLGIs. In Section III, we then introduce the quantification of disturbance created by the prior measurement enabling us to demonstrate the generic proof that WLGIs imply SLGIs. The relation between no-signaling in time conditions, LGIs and macrorealism is discussed in Section IV.  Finally, we summarize our findings in Sec.V.
\section{Standard and Wigner LGIs}
Let us consider a two-level macroscopic system at time $t_1$, which evolves from one state to another with time. According to Macrorealism \emph{per se} assumption, the system at all instant of time is found to be in a definite macroscopic state at any particular instant. Then, the measurement of a suitable dichotomic observable $\hat{M}$ should produce definite outcomes $+1$ or $-1$ . Let the measurement of $\hat{M}$ is performed on the macroscopic system at three diffrent times $t_1$, $t_2$ and $t_3 (t_3>t_2>t_1)$ which can be considered to be the measurement of the  observables $\hat{M_1}$, $\hat{M_2}$ and $\hat{M_3}$ respectively.

Now, the non-invasive measurability condition assumes that the measurement of $\hat{M_1}$ can in principle be non-invasive, so that, the definite outcome of $\hat{M_2}$ or $\hat{M_3}$ remains unaffected due to the measurement of $\hat{M_1}$ and similarly for the other set of  sequential measurements. 

 In a macrorealistic theory, by using the Macrorealism \emph{per se }and non-invasive measurability assumptions, the following inequality can then be derived,
\begin{eqnarray}
\label{lgi}
\Delta_{s}^{LG}&=&\langle M_1 M_2\rangle+\langle M_2 M_3\rangle-\langle M_1 M_3\rangle\leq1
\end{eqnarray}
which is one of the SLGIs \cite{leggett85,leggett,A.leggett} for three time measurement scenario, obeyed by a macrorealist theory. This inequality can be shown to be violated for any qubit state and maximum violation is $1.5$ \cite{leggett85}. 

 Apart from the SLGIs there have been other formulations of LGIs. One of them was derived \cite {saha} by following the Wigner formulation \cite{ep} of local realistic inequalities for stochastic hidden variable model. Wigner form of Bell's inequalities can be derived for a two-qubit entangled state of two spatially separated particles  by assuming the locality condition and existence of a global joint probabilities of the definite outcomes of the relevant dichotomic observables corresponding to the two particles. The pair-wise joint probabilities can be obtained by marginalization of the global joint probabilities. But, it is widely known that in two-party, two-input, two-output Bell scenario the only relevant inequality is the CHSH one, i.e., Wigner form of local realistic inequalities are also equivalent to the CHSH inequalities.
 
Following the Wigner idea for local realism, the stochastic inequalities(WLGIs) in a macrorealist theory  can be derived \cite{saha} by using the  statistical version of the non-invasive measurability condition that the global joint probabilities $P(M_1^{i}, M_2^{j},M_3^{k})$ (with $i,j,k=\pm$) and their marginals would remain unaffected by the measurements. For example, the joint probability $P(M_2^{+},M_3^{-})$ of obtaining the outcomes for the sequential measurements at two instants $t_2$ and $t_3$ can be obtained by marginalizations of ${M_1}$ is given by
\begin{eqnarray}
P(M_2^{+},M_3^{-})&=&\sum_{M_1=\pm}P(M_1,M_2^{+},M_3^{-})\\
\nonumber
&=&P(M_1^{+},M_2^{+},M_3^{-})+P(M_1^{-},M_2^{+},M_3^{-})
\end{eqnarray}
Writing similar other expressions for the joint probabilities $P(M_1^{+},M_2^{+})$ and $P(M_1^{-},M_3^{-})$, we get
$P(M_1^{+},M_2^{+})+P(M_1^{-},M_3^{-})-P(M_2^{+},M_3^{-})=P(M_1^{+},M_2^{+},M_3^{+})+P(M_1^{-},M_2^{-},M_3^{-})$. Invoking the non-negativity of the probability, the following form of inequality is obtained in terms of three pairs of two-time joint probabilities, is given by 
\begin{eqnarray}
\label{wlgi}
\Delta_{w}^{LG}&=&P(M_{2}^{+},M_{3}^{-})-P(M_{1}^{+},M_{2}^{+})\\
\nonumber
&-&P(M_{1}^{-},M_{3}^{-})\leq 0		
\end{eqnarray}
which is a form of WLGI, valid in a macrorealist theory.

 Note that, $23$ more such inequalities can also be derived in this manner. If we write them in the compact form, we have
\begin{eqnarray}
\label{4}
P(M_{2}^{j},M_{3}^{k})-P(M_{1}^{-i},M_{2}^{j})-P(M_{1}^{i},M_{3}^{k})\leq 0		
\end{eqnarray}
\begin{eqnarray}
P(M_{1}^{i},M_{3}^{k})-P(M_{1}^{i},M_{2}^{-j})-P(M_{2}^{j},M_{3}^{k})\leq 0
\end{eqnarray}
\begin{eqnarray}
\label{6}
P(M_{1}^{i},M_{2}^{j})-P(M_{2}^{j},M_{3}^{-k})-P(M_{1}^{i},M_{3}^{k})\leq 0
\end{eqnarray}

Now, if the LG scenario is assumed to be the temporal analogue of Bell's one, then one may expect that WLGIs are equivalent to the SLGIs. But, as mentioned earlier, there is an important fact that SLGIs do not provide necessary and sufficient condition for macrorealism in contrast to the connection between local realism and CHSH inequalities in $2222$ Bell Scenario.  While the latter represent the facets of the local realism polytope, the former does not represent the  boundaries of the macrorealism polytope\cite{clemente16}. Then, new set of inequalities may be found which can be inequivalent to the SLGIs. For a particular unsharp measurement scenario, it is numerically shown \cite{saha,pan} that the violation of WLGIs can be shown to be more robust than the violation of any of the SLGIs because the former can be violated for a larger range of sharpness parameter than the later. However, it remains unexplored if the above feature is generic.  It is the aim of the present paper to demonstrate a generic proof to show that WLGIs are inequivalent to and stronger than SLGIs.  
\section{Generic Proof to show WLGIs imply SLGIs}
In order to formulate the generic proof, we first introduce the disturbance inequalities corresponding to the various LGIs. Significance of such a disturbance inequality is that it provides the condition required for quantum violation of a relevant LGI. Note that, LGIs arise if in three time measurement scenario a restriction is imposed in the experimental arrangement by taking only the pairwise joint probabilities  (say, $P(M_2^j,M_3^k)$) into account. The pair-wise joint probabilities can also obtained by suitably marginalizing  the triple-wise (global) joint probability distribution $P(M_1^i,M_2^j,M_3^k)$ which we denote as  $P_{(M_1,M_2,M_3)}(M_2^j,M_3^k)$. However, in a macrorealist theory,  $P_{(M_1,M_2,M_3)}(M_2^j,M_3^k) $ is equivalent to $P(M_2^j,M_3^k)$. In the context of LG scenario, this is exactly the non-invasive measurability assumption at the individual level which states that a prior measurement will not change the real state of the system and its subsequent dynamics. The difference between $P_{(M_1,M_2,M_3)}(M_2^j,M_3^k)$ and $P(M_2^j,M_3^k)$ is the disturbance caused by the prior measurement at $t_1$ to the subsequent measurements at $t_2$ and $t_3$. The disturbance is the key quantity for explaining the quantum violation of a given LGI. Clearly, if no disturbance is caused by quantum measurement then all  LGIs will be satisfied.  

In order to derive the condition for the quantum violation of LGIs in terms of disturbance, let us now consider the pairwise marginal statistics of the experimental arrangement when all three measurements ($M_1$, $M_2$ and $M_3$) are performed. So that, one can write
\begin{eqnarray}
\label{triple1}
P_{(M_{1},M_{2},M_{3})}(M_{1}^{i},M_{2}^{j})&=&\sum_{k} P_{(M_{1},M_{2},M_{3})}(M_{1}^{i},M_{2}^{j},M_{3}^{k})\nonumber\\
\end{eqnarray}
\begin{eqnarray}
\label{triple3}
P_{(M_{1},M_{2},M_{3})}(M_{1}^{i},M_{3}^{k})&=&\sum_{j} P_{(M_{1},M_{2},M_{3})}(M_{1}^{i},M_{2}^{j},M_{3}^{k})\nonumber\\
\end{eqnarray}
\begin{eqnarray}
\label{triple2}
P_{(M_{1},M_{2},M_{3})}(M_{2}^{j},M_{3}^{k})&=&\sum_{i} P_{(M_{1},M_{2},M_{3})}(M_{1}^{i},M_{2}^{j},M_{3}^{k})\nonumber\\
\end{eqnarray}

where $i, j ,k=\pm$. The quantity $P_{(M_{1},M_{2},M_{3})}(M_{1}^{i},M_{3}^{k})$ conceptually differs from the pair-wise joint probability $P(M_{1}^{i},M_{2}^{j})$ where no prior measurement is performed, but in the LG context they are same by assumption of non-invasive measurability. Similar argument holds good for other pair-wise joint probabilities.

 Now, for our purpose, let us define the following quantities.
\begin{eqnarray}
\label{d11}
D_{1}(M_{2}^{j},M_{3}^{k})=P(M_{2}^{j},M_{3}^{k})-P_{(M_{1},M_{2},M_{3})}(M_{2}^{j},M_{3}^{k})\nonumber\\
\end{eqnarray}
\begin{eqnarray}
\label{d22}
D_{2}(M_{1}^{i},M_{3}^{k})=P(M_{1}^{i},M_{3}^{k})-P_{(M_{1},M_{2},M_{3})}(M_{1}^{i},M_{3}^{k})\nonumber\\
\end{eqnarray}
\begin{eqnarray}
\label{d33}
D_{3}(M_{1}^{i},M_{2}^{j})=P(M_{1}^{i},M_{2}^{j})-P_{(M_{1},M_{2},M_{3})}(M_{1}^{i},M_{2}^{j})\nonumber\\
\end{eqnarray}
Here, $D_{1}(M_{2}^{j},M_{3}^{k})$ quantifies the amount of disturbance  created (in other words, amount of violation of a no-signaling in time condition) by the measurement $M_1$ at $t_1$ to the measurements of $M_2$ and $M_3$ at $t_2$ and $t_3$ respectively. Similarly for $D_{2}(M_{1}^{i},M_{3}^{k})$. Note that, the LGIs are derived in a macrorealist model by assuming the vanishing disturbance.But in QM, the disturbance does not vanish in general. In fact, the non-zero value of the disturbance is responsible for quantum violation of LGIs. The quantity $D_{3}(M_{1}^{i},M_{2}^{j})$ quantifies the amount of disturbance created by the future measurement. Since no information can travel backward in time, $D_{3}(M_{1}^{i},M_{2}^{j})=0$ is always satisfied. We now introduce the disturbance inequalities corresponding to the various LGIs which are required to be satisfied in QM for obtaining the violation of LGIs.

Following the reasoning in \cite{maroney} and by using Eqs.(\ref{triple1}-\ref{triple2}), one can write
\begin{eqnarray}
\label{a1}
\langle M_{1}M_{2}\rangle_{(M_{1},M_{2},M_{3})}=\sum_{i,j=\pm}i j P_{(M_{1},M_{2},M_{3})}(M_{1}^{i},M_{2}^{j})\nonumber\\
\end{eqnarray}
\begin{eqnarray}
\label{a2}
\langle M_{2}M_{3}\rangle_{(M_{1},M_{2},M_{3})}=\sum_{j,k=\pm}j k P_{(M_{1},M_{2},M_{3})}(M_{2}^{j},M_{3}^{k})\nonumber\\
\end{eqnarray}
\begin{eqnarray}
\label{a3}
\langle M_{1}M_{3}\rangle_{(M_{1},M_{2},M_{3})}=\sum_{i,k=\pm}i k P_{(M_{1},M_{2},M_{3})}(M_{1}^{i},M_{3}^{k})\nonumber\\
\end{eqnarray}
Then, by using Eqs.(\ref{a1}-\ref{a3}), the expression of $\Delta_{s}^{LG}$ in ineq.(\ref{lgi}) can be written as 
\begin{eqnarray}
(\Delta_{s}^{LG})_{M_{1},M_{2},M_{3}}= \langle M_{1}M_{2}\rangle_{(M{1},M_{2},M_{3})}\nonumber\\
+\langle M_{2}M_{3}\rangle_{(M{1},M_{2},M_{3})}-\langle M_{1}M_{3}\rangle_{(M{1},M_{2},M_{3})}
\end{eqnarray}
which gives
\begin{eqnarray}
\label{d27}
(\Delta_{s}^{LG})_{M_{1},M_{2},M_{3}}=  1-4\beta
\end{eqnarray}
where $\beta =P(M_{1}^{+},M_{2}^{-},M_{3}^{+})+P(M_{1}^{-},M_{2}^{+},M_{3}^{-})$. 

Now, the difference between $\Delta^{LG}_{s}$ and $(\Delta_{s}^{LG})_{M_{1}M_{2}M_{3}}$ is the key quantity which determines whether the SLGI given by ineq.(\ref{lgi}) will be violated. Clearly, if $\Delta^{LG}_{s}=(\Delta_{s}^{LG})_{M_{1}M_{2}M_{3}}$ is satisfied, the SLGI will \textit{not} be violated. Using Eqs.(\ref{d11}) and (\ref{d22}) we can write
\begin{eqnarray}
&& \Delta^{LG}_{s}-(\Delta_{s}^{LG})_{M_{1}M_{2}M_{3}}=\sum_{j=k}D_{1}(M_{2}^{j},M_{3}^{k})\\
\nonumber&-&
\sum_{i=k}D_{2}(M_{1}^{i},M_{3}^{k})-\sum_{j\neq k} D_{1}(M_{2}^{j},M_{3}^{k})+\sum_{i\neq k}D_{2}(M_{1}^{i},M_{3}^{k})\nonumber
	\end{eqnarray}
	Since $\sum D_{1}(M_{2}^{j},M_{3}^{k})=0$, $\sum D_{2}(M_{1}^{i},M_{3}^{k}) =0$ and by noting $	\Delta^{LG}_{s}\leq 1$, we obtain
\begin{eqnarray}
	&&2\sum_{j=k}D_{1}(M_{2}^{j},M_{3}^{k})-2\sum_{i=k}D_{2}(M_{1}^{i},M_{3}^{k})\\
	\nonumber
	&+&(\Delta_{s}^{LG})_{M_{1}M_{2}M_{3}}\leq 1
	\end{eqnarray}
	By putting the value of $(\Delta_{s}^{LG})_{M_{1}M_{2}M_{3}}$ from Eq.(\ref{d27}) we have
\begin{eqnarray}
	\sum_{j=k}D_{1}(M_{2}^{j},M_{3}^{k})-\sum_{i=k}D_{2}(M_{1}^{i},M_{3}^{k})\leq 2\beta
	\end{eqnarray}
Thus, for the violation of SLGI given by ineq.(\ref{lgi}), the condition
	\begin{eqnarray}
	\label{lgsatis}
	\sum_{j=k}D_{1}(M_{2}^{j},M_{3}^{k})-\sum_{i=k}D_{2}(M_{1}^{i},M_{3}^{k}) > 2\beta
	\end{eqnarray}
needs to be satisfied in QM. We call the above relation as disturbance inequality which plays the key role in our proof. Similarly, disturbance inequalities for WLGIs can also be derived. For this, let us first consider a WLGI given by Eq.(\ref{wlgi}). Following the similar method, $\Delta_{w}^{LG}$ can be written for three measurement scenario as $(\Delta_{w}^{LG})_{M_{1}M_{2}M_{3}}=\sum_{M_{1}=\pm 1} P(M_{1},M_{2}^{+},M_{3}^{-})-\sum_{M_{3}=\pm 1} P(M_{1}^{+},M_{2}^{+},M_{3})-\sum_{M_{2}=\pm 1} P(M_{1}^{-},M_{2},M_{3}^{-})=-P(M_{1}^{+},M_{2}^{+},M_{3}^{+})-P(M_{1}^{-},M_{2}^{-},M_{3}^{-})$.

 Now, the difference between $\Delta_{w}^{LG}$ and $(\Delta_{w}^{LG})_{M_{1}M_{2}M_{3}}$ is again crucial quantity which is responsible for the violation of WLGI given by ineq.(\ref{wlgi}), and if $\Delta_{w}^{LG}=(\Delta_{w}^{LG})_{M_{1}M_{2}M_{3}}$, no violation of WLGI given by ineq.(\ref{wlgi}) can be occurred. By using Eqs.(\ref{d11}) and (\ref{d22}), we then have 
\begin{eqnarray}
\Delta_{w}^{LG}-\Delta^{LG}_{M_{1}M_{2}M_{3}}=D_{1}(M_{2}^{+},M_{3}^{-})-D_{2}(M_{1}^{-},M_{3}^{-})\nonumber\\
\end{eqnarray}
 By noting $\Delta_{w}^{LG}\leq 0$ in ineq.(\ref{wlgi}), we can then write
	 \begin{eqnarray}
	&&D_{1}(M_{2}^{+},M_{3}^{-})-D_{2}(M_{1}^{-},M_{3}^{-})\\
	\nonumber
	&-&P(M_{1}^{+},M_{2}^{+},M_{3}^{+})-P(M_{1}^{-},M_{2}^{-},M_{3}^{-})\leq 0
	\end{eqnarray}
Hence, for obtaining the violation of WLGI given by ineq.(\ref{wlgi}), the following disturbance inequality is needed to be satisfied 
	 \begin{eqnarray}
	\label{dh4v}
	&&D_{1}(M_{2}^{+},M_{3}^{-})-D_{2}(M_{1}^{-},M_{3}^{-})>\\
	&&P(M_{1}^{+},M_{2}^{+},M_{3}^{+})
	+P(M_{1}^{-},M_{2}^{-},M_{3}^{-})\nonumber
	\end{eqnarray}
Similar $23$ more disturbance inequalities (corresponding to the other $23$ WLGIs) can be derived in such a manner. If we write them in the compact notations, we have
\begin{eqnarray}
\label{dh4s}
	&&D_{1}(M_{2}^{j},M_{3}^{k})-D_{2}(M_{1}^{i},M_{3}^{k})>\\
	&& P(M_{1}^{i},M_{2}^{-j},M_{3}^{k})+P(M_{1}^{-i},M_{2}^{j},M_{3}^{-k})\nonumber
	\end{eqnarray}
	
	\begin{eqnarray}
	\label{d2}
	&&D_{2}(M_{1}^{i},M_{3}^{k})-D_{1}(M_{2}^{j},M_{3}^{k}) >\\
	&&P(M_{1}^{-i},M_{2}^{j},M_{3}^{k})	+ P(M_{1}^{i},M_{2}^{-j},M_{3}^{-k})\nonumber
	\end{eqnarray}
	
	\begin{eqnarray}
	\label{d3}
	&&-D_{2}(M_{1}^{i},M_{3}^{k})-D_{1}(M_{2}^{j},M_{3}^{-k}) >\\ 
	\nonumber
	&& P(M_{1}^{i},M_{2}^{-j},M_{3}^{k})+P(M_{1}^{-i},M_{2}^{j},M_{3}^{-k})\\
	 \nonumber
	\end{eqnarray}

Equipped with the disturbance inequalities, we are now in a position to argue that the  WLGIs imply SLGIs.  For this let us consider the following two WLGIs are given by
\begin{subequations}
\begin{eqnarray}
\label{wlgi3}
P(M_{2}^{+},M_{3 }^{+})-P(M_{1}^{-},M_{2}^{+})-P(M_{1}^{+},M_{3}^{+})\leq 0\\
\label{wlgi4}
P(M_{2}^{-},M_{3}^{-})-P(M_{1}^{-},M_{3}^{-})-P(M_{1}^{+},M_{2}^{-})\leq 0
\end{eqnarray}
\end{subequations}
 Corresponding to the WLGIs given by Eqs.(\ref{wlgi3}-\ref{wlgi4}), the relevant disturbance inequalities can be derived. They can be written as
	
	\begin{subequations}
	 \begin{eqnarray}
	\label{d111}
	D_{1}(M_{2}^{+},M_{3}^{+})-D_{2}(M_{1}^{+},M_{3}^{+})>\beta\\
	\label{d12}
	D_{1}(M_{2}^{-},M_{3}^{-})-D_{2}(M_{1}^{-},M_{3}^{-})>\beta
		\end{eqnarray}
	\end{subequations}
	For the quantum violation of the WLGIs the corresponding disturbance inequalities given by ineqs.(\ref{d111}-\ref{d12}) needs to be satisfied. Interestingly, by	adding the inequalities given by ineqs. (\ref{d111}-\ref{d12}), we get 
	\begin{eqnarray}
	\label{d15}
	\sum_{j=k}D_{1}(M_{2}^{j},M_{3}^{k})-\sum_{i=k}D_{2}(M_{1}^{i},M_{3}^{k})>2\beta
	\end{eqnarray}
	which is same as disturbance inequality in ineq.(\ref{lgsatis}) derived for SLGI.	 
	
It is now straightforward to argue that for satisfaction of ineq.(\ref{d15}), at least one of ineq.(\ref{d111}) and ineq.(\ref{d12}) needs to be independently satisfied. While satisfaction of both  ineq.(\ref{d111}) and ineq.(\ref{d12}) satisfies ineq.(\ref{d15}) but satisfaction of one of ineq.(\ref{d111}) and ineq.(\ref{d12}) does \emph{not} necessarily satisfy ineq.(\ref{d15}). Then, the satisfaction of ineq.(\ref{d15}) is the stricter condition. In other words, the violation of one of the WLGIs  does \emph{not} ensure the violation of SLGI, but converse is true.  It is crucial to note here that for any possible symmetry of SLGI, one can find two suitable WLGIs to run the similar proof presented above. We can thus write $WLGIs\Rightarrow SLGIs$.
	
	A simple example can be helpful. Consider a state in two-level system at $t_1$,
\begin{equation}
\label{state}
|\psi(t_1)\rangle = \cos\theta |0 \rangle + e^{i\phi} \sin\theta |1 \rangle
\end{equation}
with $\theta \in [0,\pi]$, $\phi \in [0,2\pi]$ and  $\hat{M_1}=\hat{\sigma_{z}}$. 
The system evolves under unitary operator $U_{\Delta t}= e^{i g \sigma_{x}\Delta t}$ in the time interval between $t_1$ and $t_2$, and $t_2$ and $t_3$. The time evolution of $M_{1}$ in two different times $t_2$ and $t_3$ are given by $M_{2}=U_{\Delta t}^{\dagger} M_{1} U_{\Delta t}$  and $M_{3}=U_{2\Delta t}^{\dagger} M_{1} U_{2\Delta t}$ respectively.\\

 At $\theta=\pi/2$ and $\phi=0$,  it can be shown that  the quantity $D_{1}(M_{2}^{j},M_{3}^{k})=0$ at $ g \Delta t=\pi/4$. But $D_{2}(M_{1}^{+},M_{3}^{+})=-1/2$, $D_{2}(M_{1}^{-},M_{3}^{-})=0$ and $\beta=1/4$. One can then find that ineq.(\ref{d111}) is satisfied, which in turn implies that WLGI given by ineq.(\ref{wlgi3}) is violated. But, ineq.(\ref{d15}) is violated which means that SLGI given by ineq.(\ref{lgi}) is satisfied. In fact, for that particular choice of state and observables none of the other symmetries of SLGI is violated.

Next, with the help of the disturbance inequalities we show how no-signaling in time condition (NSIT) implies LGIs but converse is not true.

\section{NSIT condition, LGIs and macrorealism}
  The NSIT (in other words, operational non-disturbance) condition is the statistical version of non-invasive measurability condition, seems to be analogous to the no-signaling in space condition in Bell's theorem. However, the violation of NSIT condition does not produce any fundamental inconsistency in contrast to the no-signalling in space. In the context of LG scenario,  NSIT implies that a prior measurement will not disturb the probability of obtaining an outcome of the subsequent measurements.  The satisfaction of all the NSIT conditions ( i.e., no disturbance is caused by one measurement to the other) guarantees the existence of global joint probability distribution $P(M_{1}^{i}, M_{2}^{j} , M_{3}^{k})$. 
	
A  two-time NSIT condition can then be read as 
\begin{equation}
NSIT_{(1)2}:P(M_{2}^{j})=\sum_{i} P(M^{i}_{1},M^{j}_{2})
\end{equation}
 which means that the probability $P(M_{2}^{j})$ of obtaining definite outcome is unaffected by the prior measurement of $M_{1}$ and similarly, a three-time NSIT condition is given by
\begin{eqnarray}
NSIT_{(1)23}:P(M_{2}^{j},M_{3}^{k})&=&\sum_{i}P(M_{1}^{i},M_{2}^{j},M_{3}^{k})
\end{eqnarray}
Since a future measurement cannot cause disturbance to the prior measurement, we exclude the arrow-of-time from  the present discussion.
 
In an interesting paper, Clemente and Kofler \cite{clemente} have recently  argued that  while no set of SLGIs provide the necessary and sufficient condition for macrorealism (MR), a suitable conjunction two-time and three-time NSIT conditions provides the same. They argued that 
\begin{eqnarray}
\label{nsit}
NSIT_{(2)3} \wedge NSIT_{(1)23} \wedge NSIT_{1(2)3} \Leftrightarrow MR
\end{eqnarray}
The choice of two-time NSIT condition is  not unique. One may replace  $NSIT_{(2)3}$ by $NSIT_{(1)3}$. 

In terms of the notion of disturbance is defined in Eqs.(\ref{d11}-\ref{d33}), the above Eq.(\ref{nsit}) can be recast as

\begin{eqnarray}
\label{nsit1}
D_{1,2}(M_{3}^{k})\wedge D_{1}(M_{2}^{j},M_{3}^{k})\wedge D_{2}(M_{1}^{i},M_{3}^{k})  \Leftrightarrow MR
\end{eqnarray}

Let us now closely examine the connection between NSIT condition, LGIs and macrorealism in the context of LG measurement scenario considered here. Note again here that the violation of NSIT implies that a prior measurement caused a detectable disturbance to subsequent measurements. 

It can then be seen from the ineq.(\ref{d15}) that non-zero value of either or both of $D_{1}(M_{2}^{j},M_{3}^{k})$ and $D_{2}(M_{1}^{i},M_{3}^{k})$ are necessary condition for SLGI. However, this is not sufficient. In order to obtain the violation of SLGI the interplay between the  disturbances plays crucial role so that the difference between two disturbances is greater than $2\beta$, where $\beta$ is a positive quantity.
To make the argument simpler, let for a specific choice of state and observable, $D_{1}(M_{2}^{j},M_{3}^{k})\neq 0$ but $D_{2}(M_{1}^{i},M_{3}^{k})=0$. This does not ensure the satisfaction of the ineq.(\ref{d15})  which requires $\sum_{j=k}D_{1}(M_{2}^{j},M_{3}^{k})>2\beta$. In other words, the mere violation of NSIT is \emph{not} enough to ensure the violation of a LGI, unless the amount of violation is larger than a \emph{threshold value} of the disturbance. Similar arguments for the violation of WLGIs can be drawn from Eqs.(\ref{d111}) and (\ref{d12}). 

 Then, by following the argument in \cite{clemente16}, we can say that the violation of one of the NSIT conditions ensures the violation of macrorealism but not WLGIs or SLGIs. We thus explained how NSIT is a better candidate for testing macrorealism by using the disturbance inequalities. 
\section{Summary and Discussions}
In this paper, for a three-time LG measurement scenario, we examined the inequivalence between various formulations of Leggett-Garg inequalities, viz., standard Leggett-Garg Inequalities (SLGIs), Wigner form of Leggett-Garg Inequalities (WLGIs) and no-signaling in time (NSIT) conditions.  For $2222$ Bell scenario, the only relevant Bell's inequality is the CHSH form. Any stochastic version, such as, CH and Wigner form of Bell's inequalities are equivalent to the CHSH inequalities in $2222$  Bell Scenario. We showed here that although SLGIs seem to be a temporal analogue of CHSH inequalities, but WLGIs are \emph{ inequivalent} to the SLGIs. We provided a generic proof to demonstrate  that WLGIs are stronger than SLGIs, in the sense that the quantum violation of any of the three possible symmetries of SLGIs ensures the violation of at least one of the twenty four WLGIs. But, the converse does \textit{not} hold good. Thus, WLGIs provide better test of macrorealism than SLGIs - a fact, which seems amenable to experimentally test the notion of macrorealism.  Importantly, our proof is independent of dimension of the system and irrespective of any particular measurement scheme. This is done by introducing a measure of the amount of disturbance caused by the prior measurement to the subsequent measurements and then by deriving disturbance inequalities corresponding to the various LGIs. 
	
In a recent work \cite{clemente}, it is shown that a suitable conjunction of NSIT conditions provide the necessary and sufficient condition for macrorealism while LGIs cannot provide the same. One can then argue that NSIT conditions captures the notion of macrorealism better than LGIs. In terms of disturbance inequality,  here we demonstrated how NSIT conditions provide a better test of macrorealism. However, such a proof can not be demonstrated for local realistic inequalities due to the fact that the statistically detected disturbance caused by space-like separated events is restricted by no-signalling in space condition.

Recently, Clemente and Kofler \cite{clemente16} have proved that no set of SLGIs can provide necessary and sufficient condition for macrorealism - a feature which is in contrast to the Fine's theorem\cite{fine} for local realism. By referring to our earlier discussion, we can also argue that no set of WLGIs can provide the necessary and sufficient condition for macrorealism. In view of our study it would be instructive to formulate new set of inequalities to examine whether that new set along with the existing set of LGIs provides the same for macrorealism. A study along this direction is very recently initiated \cite{hall} by using a quasi-probability approach. A comparison is also made\cite{hall17}  between SLGI and NSIT conditions by introducing the notion of weak and strong form of macrorealism. It is argued that two-time LGIs provide the necessary and sufficient condition for weak form of macrorealism. Further studies are required to analyse this result. We note the fact that for testing local realism the joint measurement needs to be considered between separated system. In contrast, the joint measurement in LG scenario is more flexible thereby indicating the possibility of constructing new set of inequalities differing from WLGIs.  Since three-time SLGIs and WLGIs are not the facets of macrorealism polytope, it would then be interesting to examine whether such a new set of inequalities provide the necessary and sufficient condition for macrorealism. Studies along this direction could be an interesting avenue of future research.
    
\acknowledgments
AKP acknowledges the support from Ramanujan Fellowship research grant (SB/S2/RJN-083/2014). 


\begin{thebibliography}{0}
\bibitem{bell}J. S. Bell, Physics, 1, 195 (1964).
\bibitem{sch}E. Schroedinger, Naturwissenschaften, 23, 807 (1935).
\bibitem{zeh}H. D. Zeh,  Found. Phys. 1, 69 (1970); W. H. Zurek, Phys. Rev. D 26, 1862 (1982).
\bibitem{ghi} G. C. Ghirardi, A. Rimini, and T. Weber, Phys. Rev. D 34, 470 (1986); A. Bassi and G. Ghirardi, Phys. Rep. 379, 257 (2003).
\bibitem{bruk}J. Kofler and C. Brukner,Phys. Rev. Lett. 99,
180403 (2007).
\bibitem{arndt} M. Arndt \textit{et al}., Nature, 401, 680 (1999); S. Gerlich \textit{et al}., Nature Comm. 2, 263 (2011).
\bibitem{leggett85}A. J. Leggett and A. Garg, Phys. Rev. Lett. 54, 857 
(1985).
\bibitem{leggett} A. J. Leggett, J. Phys. Condens. Matter. 14, R415 (2002).
\bibitem{A.leggett} A. J. Leggett, Rep. Prog. Phys. 71, 022001 (2008).
\bibitem{kofler}J. Kofler and C. Brukner, Phys. Rev. A 87, 052115
(2013).
\bibitem{saha} D. Saha, S. Mal, P. K. Panigrahi and D. Home, Phys. Rev. A, 91, 032117 (2015).
\bibitem{clemente}L. Clemente and J. Kofler, Phys. Rev. A, 91, 062103 (2015).
\bibitem{clemente16}L. Clemente and J. Kofler, Phys. Rev. Lett. 116, 150401 (2016).
\bibitem{UshaDevi} A. R. Usha Devi, H. S. Kartik, Sudha and A. K. Rajagopal,
 Phys. Rev.A, 87, 052103 (2013).
\bibitem{emary} C. Emary, N. Lambert and F. Nori, Rep. Prog. Phys. 77, 016001 (2014).
\bibitem{maroney} O. J. E Maroney and C. G Timpson, arXiv:1412.6139.
\bibitem{budroni}C. Budroni and C. Emary, Phys. Rev. Lett., 113, 050401 (2014).
\bibitem{budroni15} C. Budroni \emph{et al.,} Phys. Rev. Lett., 115, 200403 (2015).
\bibitem{moreira}S. V. Moreira, A. Keller, T. Coudreau and P. Milman, Phys. Rev. A, 92, 062132 (2015).
\bibitem{hall} J. J. Halliwell, Phys. Rev. A, 93, 022123 (2016).
\bibitem{pan} S. Kumari and A. K. Pan, arXiv:1705.09934.
\bibitem{mal} S. Mal and A. S. Majumdar, Phys. Lett. A, 380, 2265(2016).
\bibitem{a.p} A. Palacios-Laloy \textit{et al}., Nat. Phys. 6, 442 (2010).
\bibitem{goggin} M. E. Goggin \textit{et al}., Proc. Natl. Acad. Sci. USA 108,
1256 (2011).
\bibitem{xu} J.-S. Xu, C.-F. Li, X.-B. Zou, and G.C. Guo, Sci. Rep. 1, 101 (2011).
\bibitem{dressel}J. Dressel, C. J. Broadbent, J. C. Howell, and A. N. Jordan, Phys. Rev. Lett. 106, 040402 (2011).
\bibitem{suzuki} Y. Suzuki, M. Iinuma, and H. F. Hofmann, New J. Phys. 14, 103022 (2012).
\bibitem{julsgaard} B. Julsgaard, A. Kozhekin, and E. S. Polzik, Nature, 413, 400 (2001).
\bibitem{isart} O. Romero-Isart \textit{et al}., Phys. Rev. Lett. 107, 020405 (2011).
\bibitem{mahesh11} V. Athalye, S. S. Roy and T. S. Mahesh, Phys. Rev. Lett. 107, 130402 (2011). 
\bibitem{souza} A. M. Souza, I. S. Oliveira, and R. S. Sarthour, New J. Phys. 13, 053023 (2011).
\bibitem{katiyar13} H. Katiyar, A. Shukla, K. R. K. Rao and T. S. Mahesh,
Phys. Rev. A, 87, 052102(2013).
\bibitem{formaggio} J. A. Formaggio, D. I. Kaiser, M. M. Murskyj and T. E. Weiss,
Phys. Rev. Lett. 117, 050402 (2016).
\bibitem{katiyar} H. Katiyar, A. Brodutch, D. Lu, R. Laflamme,
New J. Phys. 19, 023023 (2017).
\bibitem{fine} A. Fine, Phys. Rev. Lett. 48, 291(1982).
\bibitem{chsh}  J. F. Clauser, M. A. Horne, A. Shimony and R. A. Holt,
Phys. Rev. Lett. 23, 880 (1969).
\bibitem{col}D. Collins and N. Gisin, J. Phys. A: Math. Gen., 37, 1775 (2004).
\bibitem{cglmp} D. Collins, N. Gisin, N. Linden, S. Massar and S. Popescu, Phys. Rev. Lett. 88, 040404 (2002).
\bibitem{ch} J. F. Clauser, M. A. Horne, Phys. Rev. D 10, 526 (1974).
\bibitem{ep} E. P. Wigner, Am. J. Phys. 38, 1005 (1970).
\bibitem{cas} S. Castelletto, I. P. Degiovanni, and M. L. Rastello, Phys. Rev. A 67, 044303 (2003).
\bibitem{home15}D. Home, D. Saha and S. Das, Phys. Rev.A, 91, 012102 (2015).
\bibitem{hall17} J. J. Halliwell, Phys. Rev. A 96, 012121 (2017).
\end{thebibliography}
\end{document}